\documentstyle[prl,multicol,twocolumn,aps]{revtex}

\begin{document}

\onecolumn
\draft
\title{Dynamics of Mesoscopic Precipitate Lattices
 in Phase Separating Alloys under External Load}
\author{R.~Weinkamer$^{1,2,3}$,
        H.~Gupta$^{2,3}$,
        P.~Fratzl$^3$,
        and J.L.~Lebowitz$^2$}
\address{$^1$Institut f\"ur Materialphysik der Universit\"at Wien,
  Strudlhofgasse 4, A-1090 Wien, Austria\\
  $^2$Departments of Mathematics and Physics, Rutgers University,
         Busch Campus, New Brunswick, 08903 New Jersey, USA\\
         $^3$Erich Schmid Institut f\"ur Materialwissenschaft,
         \"Osterreichische Akademie der Wissenschaften
         \& Institut f\"ur Metallphysik, Montanuniversit\"at Leoben,
         Jahnstra\ss e 12, A-8700 Leoben, Austria}

\maketitle

\begin{abstract}
We investigate, via three-dimensional atomistic computer simulations, phase
separation in an alloy under external load. A regular two-dimensional array
of cylindrical precipitates, forming a mesoscopic precipitate lattice, 
evolves in the case of applied tensile stress by the movement of mesoscopic
lattice defects. A striking similarity to ordinary crystals is found in the
movement of "meso-dislocations", but new mechanisms are also
observed. Point defects such as "meso-vacancies" or "meso-interstitials"
are created or annihilated locally by merging and splitting of
precipitates. When the system is subjected to compressive stress, we
observe stacking faults in the mesoscopic one-dimensional array of
plate-like precipitates.

\end{abstract}

The coarsening of precipitates in a phase separating alloy usually proceeds
via an Ostwald ripening process where large particles grow at the expense
of smaller ones \cite{wagnerk}. The driving force for this process is the
tendency to reduce the total amount of interface between matrix and
precipitates.  The coarsening kinetics follows the laws of dynamical
scaling, with a characteristic length $R(t)$ growing like $t^{1/3}$, as
first described by the theory of Lifshitz, Slyozov and Wagner (LSW theory)
when the
density of precipitate particles is small \cite {lifshitz,alkemper}.
The situation changes radically when a difference in lattice spacing
between matrix and precipitates (lattice misfit) gives rise to long-range
elastic interactions. The precipitates then tend to be cuboidal or
plate-like instead of spherical and start to align themselves into more or
less regular arrays to minimize the elastic energy 
\cite {fratzl1,khach1,thompson,su,orlikowski}.
Changes in the growth kinetics of precipitates,
especially a slowing-down in the later stages, have also been reported
\cite{miyazaki,onuki}.  When a uniaxial external stress is applied, the cubic
symmetry is broken and precipitates arrange themselves into arrays of
cylinders parallel to or of plates perpendicular to the
direction of external stress (rafting) \cite{chang,laberge,hort,lee,li1,li}.

During the coarsening process in an alloy with lattice misfit,
the number of precipitates reduces, implying an increase in the 
spacing between them. Since this cannot occur continuously in a regular
array of precipitates,
Wang et al. \cite{wang} have proposed
defects within the mesoscopic lattice of precipitates to allow
the necessary rearrangement in a way very similar to dislocation climbs in
ordinary crystals.  Nothing is known, however, about the
detailed mechanisms that allow the generation and the movement of
such mesoscopic defects in order to enable the coarsening process.
The dynamics of these defects rather than the atomic diffusion alone,
as in the LSW theory, will have a decisive effect on the growth
kinetics of the precipitates. Similar phenomena may be expected
in the
coarsening of the two-dimensional arrays of cylinders or the
one-dimensional array of plates occurring as a consequence of rafting
under uniaxial stress.

To clarify the mechanisms by which the precipitate lattices
coarsen, it is necessary to investigate the details of the
local arrangements of the precipitates and their movement, in the same
way as it is necessary to study local atomic arrangements for the
detailed investigation of dislocation movements at the atomic level.
While the latter is possible with transmission electron microscopy,
the first is much more difficult if not impossible with direct
imaging or scattering techniques.
In the present paper, we
study this phenomenon by three-dimensional
computer simulations using an atomistic lattice model that has proven
effective in the simulation of phase separation with lattice misfit
in two dimensions \cite{fratzl2,laberge}.
One of the advantages of such  computer simulations is that
the process can be followed "in-situ" by monitoring the evolution of
individual precipitates. An investigation of phase
separation under external load has the further advantage that 
the dimensionality of the precipitate lattices is lowered due to the 
external stress (two-dimensional for tensile, one-dimensional for 
compressive stress, respectively) making an analysis of the coarsening
process easier.
In the case of cylindrical precipitates parallel to the direction of tensile
stress, it turns out that the "meso-lattice" formed by the
projected cylinder axes
evolves by the movement of lattice defects.
We want to emphasize the similarity of their appearance to atomistic
lattice defects by naming them
"meso-dislocations", "meso-vacancies" and "meso-interstitials".
Whereas meso-dislocations move in a way very similar to dislocations in an
ordinary crystal, the dynamics of meso-vacancies and
-interstitials is totally
different from their analog in atomic lattices. There is no long-range
diffusion of these defects, but they are created or annihilated locally
by splitting or merging of cylindrical precipitates.
In the case of an applied compressive stress, we show the existence of
"meso-stacking faults" in the
linear array of plates perpendicular to the direction of stress.

We use a conceptually simple three-dimensional atomistic model for a binary
alloy
containing only parameters which are experimentally accessible.
A face-centered cubic (fcc) lattice ${\cal L}$
with  $N$ lattice sites and periodic boundary conditions is
occupied by two types of atoms, $A$ and $B$.
A spin
variable $\gamma({\bf p})$ is assigned at each ${\bf p} \epsilon {\cal
L}$, with $\gamma({\bf p})=1$ if there is an $A$ atom at site ${\bf p}$
and $\gamma({\bf p})=-1$ if there is a $B$ atom at this site.
A chemical interaction between nearest-neighbor atoms is assumed with like
atoms attracting one another.
To model the elastic
interactions caused by the different sizes of $A$ and $B$ atoms,
all nearest-neighbor pairs of atoms are considered to be
connected by springs with a longitudinal and two different transverse spring
constants.
The Hamiltonian of this microelastic model can be written as a quadratic
form
of the spin variables $\gamma({\bf p})$ or, using their Fourier transforms
$\tilde{\gamma}({\bf k})$, as
\cite {khach,cook,khach1,fratzl2,laberge}

\begin{equation}
{\cal H} = \frac{1}{2N}\sum_{{\bf
k}}{\bf \Psi}({\bf k})\mid\tilde{\gamma}({\bf k})\mid^{2},
\end{equation}
when it is assumed that the spring constants are independent of composition
and the relaxation time of the lattice distortions is much shorter than the
diffusion time of the atoms.  In the case of external applied stresses and
only a weak dependence of the spring constants on the type of atoms they
connect, the expression (1) remains approximately valid but the interaction
potential ${\bf \Psi}({\bf k})$ does not have same symmetry as the crystal
lattice (for details on the model see \cite{laberge}).

The atomic configurations on the lattice evolve by
the Metropolis algorithm with
Kawasaki exchange dynamics of nearest-neighbor atoms \cite {binder}.
To deal with the
long-range interactions a special updating procedure was employed
\cite{fratzl2}.
The time unit is one Monte Carlo step (MCS), i.e. one attempted exchange
per site.
We carried out  simulations on a lattice ${\cal L}$ with 48 cubic cells in
each direction, i.e. 442368 lattice sites.
The size difference between A and B atoms was taken to be
1\%.
The spring constants in the model correspond to
Born-von Karman parameters, which can be obtained
directly from the measured phonon dispersion of the
crystal \cite{landolt}.
Choosing the parameters of copper, our model
resembles elastically a typical metal with negative elastic
anisotropy.
The concentration
of A atoms was taken $\bar{c}=0.2$. Starting from a homogeneous
configuration,
the time
evolution of the system was monitored after a quench to a temperature
within the two-phase region
$T /T_{c} \approx 0.4$, where $T_c$ denotes the critical temperature
for phase separation determined approximately by a series of MC runs.
Finally, the sign of the misfit was chosen in such a way that
cylindrical precipitates form under tensile stress and plates
under compressive stress \cite{chang,laberge}.

In Figure 1 the mean precipitate size $R(t)$ is plotted as a function of
$t^{1/3}$. $R(t)$ is defined as $R(t)\equiv \frac{2\bar{c}}{P_{AB}}$, where
$P_{AB}$ is the probability of a nearest-neighbor $AB$ atom-pair. This 
definition assumes that during coarsening $R(t)$ is inversely proportional
to the interface between precipitate and matrix, which is estimated in
our model by the probability $P_{AB}$. Besides the symbols
representing the results for the cases 
of tensile, compressive and no external stress, the solid line depicts 
$R(t)$ for an $fcc$ Ising model with only nearest neighbor chemical 
interactions at the same temperature compared to $T_c$. After ~ 600 MCS
the growth of
the precipitates slows down substantially from the usual power law
$R(t)\propto t^{1/3}$ in the case when elastic interactions are considered,
although $R(t)$ is nearly insensitive to the sign of the external load.
An influence on $R(t)$ due to the finite system size cannot be fully excluded.
A study of a system with comparable system size (but with short range chemical 
interactions only) did not show any finite-size
effects at the comparatively early times ($t < 6000$ MCS) we consider here \cite{heermann}.
 
To understand the processes leading to this slowing down, we looked
in detail at the time evolution of the precipitate microstructure.
Figure 2 shows the configuration obtained after 3000 MCS when the system was
subjected to tensile stress. To reduce the three-dimensional configuration
to
a two-dimensional plot, we averaged the concentration of $A$ atoms along
lines normal to the plotting plane. Therefore the plot
can be read in a way similar to an X-ray transmission micrograph.
An analysis of the full three-dimensional data as well as
the structure function confirmed that
the more or less regularly arranged bright spots in
Fig. 2 correspond to $A$-rich, cylindrical precipitates with their
axis parallel to the stress direction.
For a more detailed analysis of the kinetics, we have redrawn these
"micrographs" using contour lines to indicate the amount of $A$ atoms
along a line perpendicular to the plot (Fig. 3). For a clearer
representation of the data, they were smoothed
by averaging the value of each pixel with all of its first
neighbors. Peaks in the amount of $A$ atoms are visible in the contour plots
and can be interpreted as points of a "meso-lattice" of parallel cylindrical
precipitates. These points were connected by lines revealing a distorted
square lattice reflecting the symmetry of the
elastically soft $\langle100\rangle$-directions.
Contour lines were plotted for steps of 0.1 in the
averaged concentration
of $A$ atoms starting at $c= \bar{c} = 0.2$.
The highest values around the "meso-lattice points"
correspond to $c \geq 0.9$. In some cases, however, the
height is significantly smaller (e.g., the point just below
the number 4 in Fig. 3a).  These correspond to cylinders which do 
not extend over the full thickness of the specimen along
the line of projection.
Tilted or kinked cylinders may give elongated spots in the
contour line plot (e.g., close to number 8 and 11 in Fig. 3c).

The number of cylindrical precipitates decreases from 72 at 600 MCS
to 49 at 3000 MCS. The spacing of the
mesoscopic lattice increases visibly during this period.
A measure for this increase is the number of rows and columns
as given by the square root of the number
of precipitates. Starting with 8.48
(600 MCS),
the number decreases continuously
to 8.19 (1200 MCS), 7.35 (2000 MCS)
and finally 7 (3000 MCS). Such a continuous change via non-integer
numbers is, however, impossible in a perfect square lattice. Indeed,
Fig. 3 shows defects responsible for this increase in the
meso-lattice parameter via the following dynamical processes:

1) {\it Dynamics of meso-interstitials}: At 600 MCS
we observe a number of precipitates which do not occupy
regular lattice sites (marked
by open circles in Fig. 3). We call these meso-interstitials.
The contours show that meso-interstitials dissolve by
a flow of $A$ atoms along elastically soft directions
towards a
nearest-neighbor precipitate on a regular lattice site.
In some cases this also leads to a shift of the defect
which finally merges with the precipitate of the
meso-lattice site (arrow in Fig. 3a,b). Most of these
meso-interstitials are not fully developed cylinders as
indicated by their relatively low content in $A$ atoms
as compared to regular lattice sites.
During the period covered by Fig. 3,
new meso-interstitials (Fig. 3c) are not formed by
precipitate nucleation, but
by rearrangement of the meso-lattice (see point 3 below).

2) {\it Dynamics of meso-vacancies}:
Splitting of precipitate cylinders is also possible
(arrow in Fig. 3d).
In the case shown, a precipitate splits to fill
a neighboring meso-vacancy (number 11 in Fig. 3c).
Meso-vacancies are represented by squares in Fig. 3.
They typically are formed by dissolution of precipitates
on regular meso-lattice sites, which is one of the mechanisms
to reduce the total number of lattice points
(see vacancies 9 to 11 in Fig. 3c,d).

3) {\it Dynamics of meso-dislocations}:
In addition to the already mentioned vacancies and interstitials, the
meso-lattice also shows the analog of dislocations (denoted by the symbol
$\perp$). They interact with point defects and also with each other to
increase
the spacing and regularity of the meso-lattice. Examples for such complex
reactions are visible in Fig. 3.  First, the column of points bounded by
$\perp 1$
and $\perp 4$ is gradually disappearing from Fig. 3a to 3c by dissolution
of precipitates leading to the annihilation of the meso-dislocations.
The start of a similar process can be seen with the formation of vacancy 9
and the subsequent dissolution of the two  lattice points immediately below
(which
has almost led to their disappearance in Fig. 3d).
Another 
process starts with the dissolution of a single lattice point 
(denoted by $\star $
in Fig. 3a). The resulting meso-vacancy splits into two dislocations 
(denoted $\perp 7$ and $\perp 8$ in Fig. 3b). 
$\perp 7$ is attracted by $\perp 3$ and 
they finally 
annihilate leaving behind two interstitials (Fig. 3c) which both dissolve. 
Only
the remains of one of these two are still visible in Fig. 3d.

When the system is subjected to compressive stress, our
simulations show a one-dimensional array of plate-like 
precipitates corresponding to regions of high concentration of A atoms perpendicular
to the direction of applied stress. 
 These plate like precipitates are separated by regions with hardly
any A atoms at all.  Fig. 4 shows on top the
one-dimensional concentration profile of $A$ atoms 
{\it along} 
the stress direction.
Near the middle of the system there are large defects almost washing out the
periodic structure. But even where the plates seem
well-defined,
lattice defects can be observed. This is illustrated by the three
pictures at the bottom which give the distribution of $A$ atoms
within
planes {\it perpendicular} to the direction of stress. 
Holes in the plate-like precipitates (plate 1 and plate 2 in Fig. 4) 
are aligned along the elastically
soft cubic directions.
The most striking feature, however, is the defect highlighted by the dashed
rectangle in all three pictures.
In this region, the plates have a large hole, while in the gap between the plates
(gap in Fig. 4)
an $A$-rich area can be seen. 
This highlighted region with inverted content of $A$-atoms 
therefore corresponds
to a stacking fault of the meso-lattice.
A schematic view of the defect (with the components plate 1 , gap and plate 2)
 is also shown in Fig. 4. 
Such stacking faults occur
frequently and are extremely
long-lived in the time-evolution of the alloy, particularly at low
temperatures.  

The occurrence of defects in mesoscopic precipitate lattices with totally
different topology suggests that they are a common feature when coherent
elastic strains are important in the phase transformation process.
We expect that similar defects can be found in a system, which is
not subjected to an additional external stress.
The simplified topology due to the external stress helped to
identify several types of
defects, showing similarities and some striking
differences with respect to defects in ordinary crystal lattices.
In contrast to the usual Ostwald ripening process,
coarsening is mediated by the movement of these mesoscopic defects,
which allows a reduction of the number of precipitates without
destroying the mesoscopic lattice structure.

\acknowledgments
The work of H.G. and J.L.L. was supported by NSF Grant No. NSF-DMR-9813268.

\begin{figure}
\caption{ 
Time evolution of the mean precipitate size $R(t)$ on a $t^{1/3}$ scale 
for the case of applied tensile (squares), compressive (circles) and 
no external stress (triangles) at a temperature $T/T_c \approx 0.4$.
The solid line corresponds to an Ising model with nearest neighbor
chemical interactions only. 
} \label{mean size}
\end{figure}

\begin{figure}
\caption{ 
Concentration of solute $A$ atoms after 3000 MCS averaged along the 
direction of tensile stress, i.e., normal to the plotting plane.
Bright (dark) regions represent a high (low)
concentration of A atoms.
} \label{average}
\end{figure}
 
\begin{figure} 
\caption{  
Time evolution of the two-dimensional array of cylindrical precipitates (black
points) formed under tensile stress: a) 600 MCS, b) 1200 MCS, c) 2000 MCS and
d) 3000 MCS. Lattice defects in the 
mesoscopic precipitate lattice, such as meso-dislocations (denoted
by $\perp$),
meso-vacancies ($\Box$) and meso-interstitials ($\bigcirc$) are
marked and labeled. The underlying contour plots provide information about the 
concentration of $A$ atoms averaged along the direction of the tensile stress. 
} \label{dislocations}
\end{figure}

\begin{figure} 
\caption{
One-dimensional concentration profile of $A$ atoms (graph at the top) along 
the direction of 
compressive stress after 3000 MCS.
The three pictures at the bottom show the concentration of $A$ 
atoms normal to
the direction of stress after averaging over 
several neighboring lattice planes; a total of 6 lattice planes for
the plate-like precipitates (plates 1 and 2) and 8 lattice planes 
for the gap in-between. Regions of high $A$ atom concentration are bright.
The dashed line highlights a mesoscopic 
stacking fault in the one-dimensional array of plate-like precipitates.
A three dimensional schematic view of the defect is given in the middle of the 
figure, where the arrows mark the direction of the external stress.  
} \label{stacking}
\end{figure} 
 
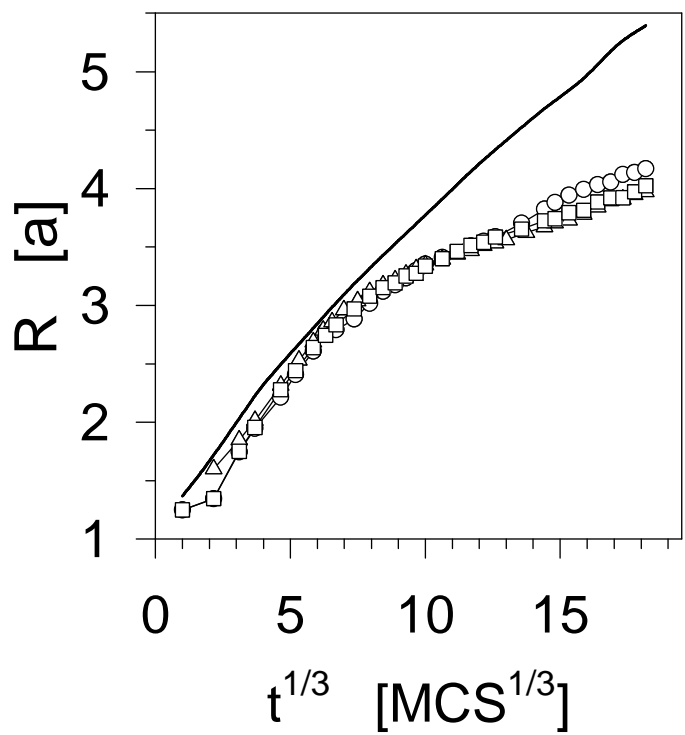
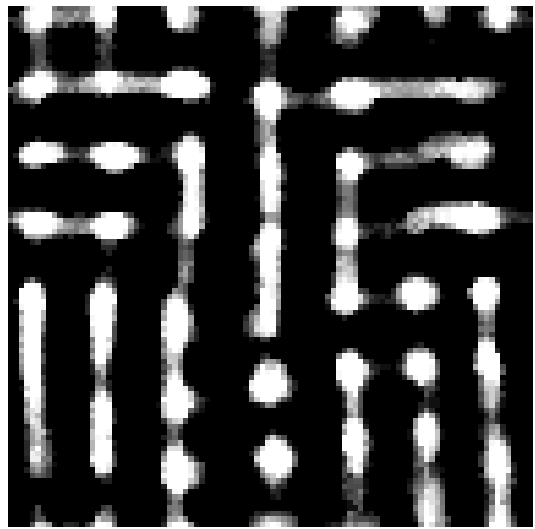
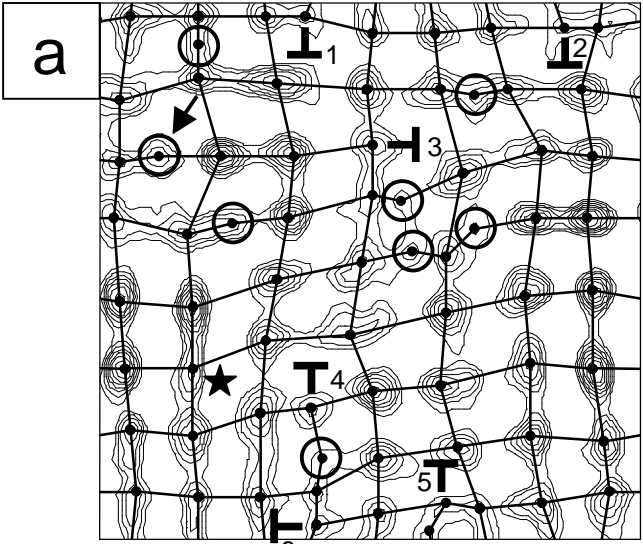
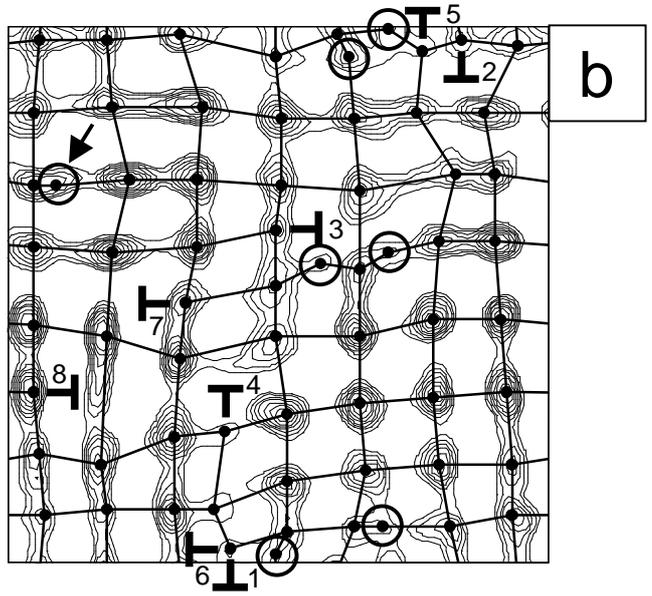
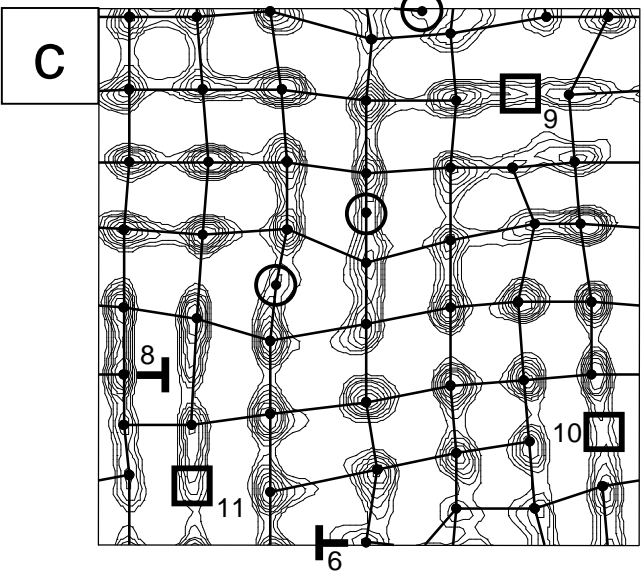
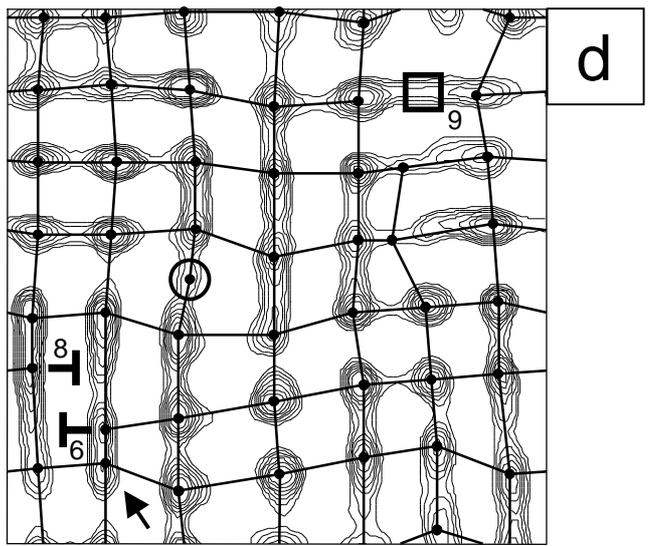
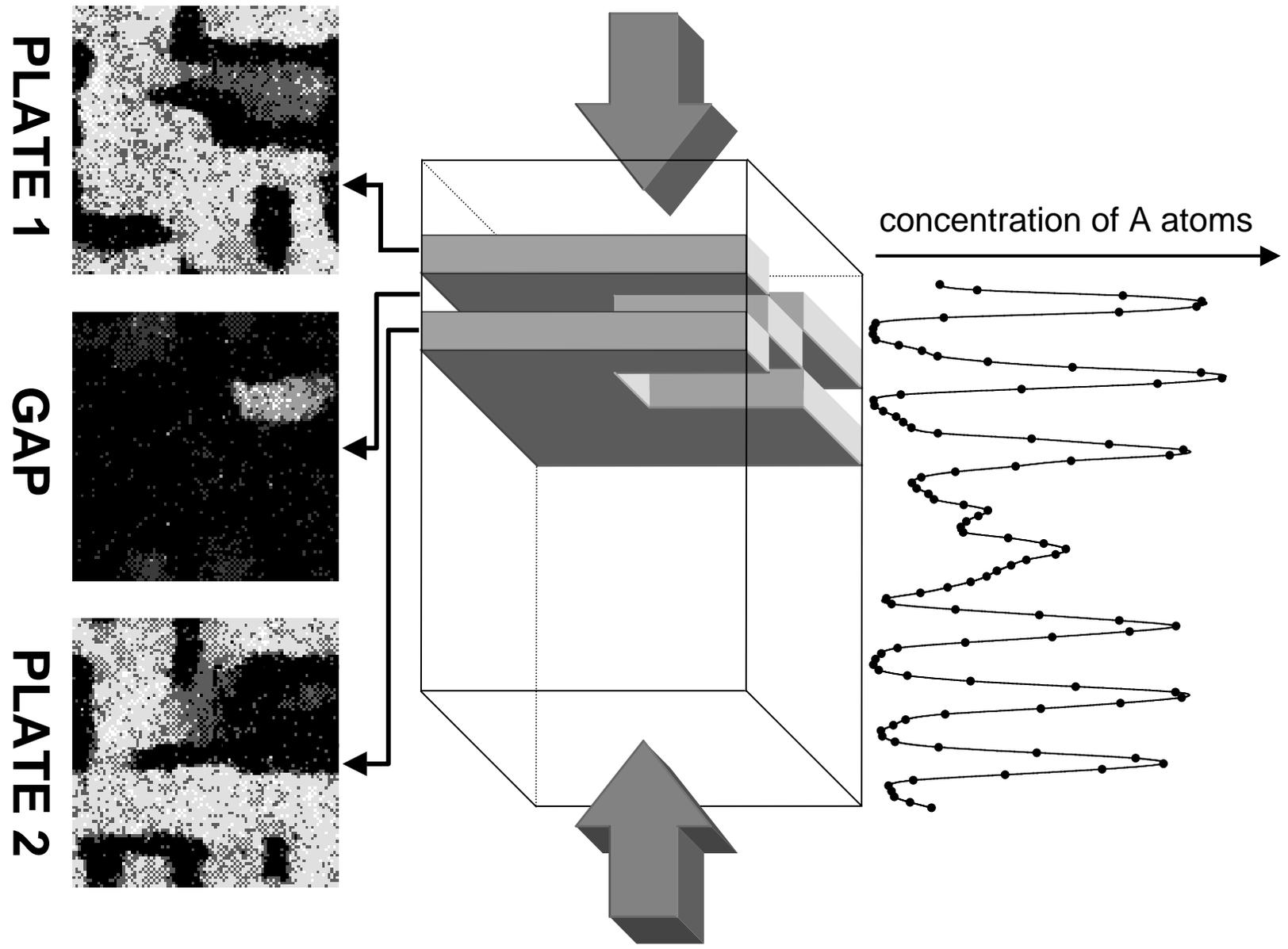
\end{document}